# Evidence for First Order Cubic Paraelectric to Rhombohedral Ferroelectric Phase Transition in $0.8BiFeO_3$-$0.2Pb(Fe_{1/2}Nb_{1/2})O_3$

Jay Prakash Patel, Anar Singh and Dhananjai Pandey[*]

School of Materials Science and Technology, Institute of Technology, Banaras Hindu University, Varanasi-221005, India

The current controversies about the existence of an intermediate 'β' phase of $BiFeO_3$ and the high temperature paraelectric 'γ' phase are resolved by studying the sequence of ferroic transitions in $0.8BiFeO_3$-$0.2Pb(Fe_{1/2}Nb_{1/2})O_3$ (BF-0.2PFN) with lowered transition temperature. It is shown that the room temperature ferroelectric phase of 0.8BF-0.2PFN in the R3c space group transforms to the paraelectric/paraelastic cubic ($Pm\bar{3}m$) phase directly without any intermediate 'β' phase reported in the literature. This transition is of first order type as confirmed by the coexistence of R3c and $Pm\bar{3}m$ phases over a 100K range and a discontinuous change in the unit cell volume.

[*] Author to whom correspondence should be addressed;
electronic mail: dpandey_bhu@yahoo.co.in

The current interest in the magnetoelectric multiferroic materials [1-3] is driven by the possibility of developing new generation actuator, sensor, and storage devices [3-5]. The coexistence and coupling of magnetic and ferroelectric orders in $ABO_3$ type perovskites is unusual from the point of view of the underlying physics behind the two phenomena, as the former requires partially filled 'd' orbitals of the B site cation while the latter is driven by the hybridization of empty 'd' oribitals of B atom with partially filled oxygen '2p' orbitals [6, 7]. Last few years have witnessed several theoretical efforts to understand the coexistence and coupling of these two mutually exclusively phenomenona [2, 6, 8, 9].

$BiFeO_3$ is the only room temperature magnetoelectric multiferroic. It has got the highest ferroelectric polarization (50 to $100\mu C/cm^2$ measured on insulating single crystals) [10, 11], ferroelectric Curie temperature ($T_C$ ~1103K) [12] and Neel temperature ($T_N$ ~643K) [13] for G type antiferromagnetism with an incommensurate cycloidal magnetic ordering in the $[110]_h$ direction [14]. $BiFeO_3$ has therefore evoked considerable attention from the point of view of coexistence of ferroelectric and magnetic orders, and their magnetoelectric coupling, over the last several decades [4, 5]. It has been shown in recent years that ferroelectricity in this compound arises due to lone pair stereochemistry of Bi 6s orbitals [6] and not due to the covalency effects of Fe-O bonds [7]. Recent work of Singh et al on a $BiFeO_3$ based system has provided direct atomic level evidence for magnetoelectric coupling of intrinsic multiferroic origin in terms of an isostructural phase transition accompanying the development of magnetic order leading to significant shifts of atomic positions and excess polarization [15]. Magnetoelectric coupling of intrinsic multiferroic origin has also been demonstrated in single crystals through switching of the direction of the cycloidal magnetic order by electric field [16, 17].

While there has been considerable progress in the understanding of the magnetoelectric coupling and the magnetic structure of $BiFeO_3$, the stability field of its room temperature phase and the sequence of the structural phase transitions it undergoes as a function of temperature continue to remain controversial and has a long history too, as discussed in the recent literature [18-23]. There is unanimity on the room temperature crystal structure of the so-called α-phase of $BiFeO_3$. It corresponds to a rhombohedrally distorted pervoskite structure in the R3c space group and $a^-a^-a^-$ (in Glazer's notation [24]) tilt system. The controversial issues currently under intense debate are [18-23]: (i) Is there a β-phase of $BiFeO_3$ to which the room temperature α-phase in the R3c space group transforms at the ferroelectric/ferroelastic $T_C$? (ii) What is the space group of this β-phase? (iii) Is there an unambiguous evidence for the existence of the paraelectric/paraelastic cubic phase (γ-phase) prior to melting /decomposition of $BiFeO_3$? The controversy about the structure of the β-phase can be assessed from the fact that as many as five different space groups, I4/mcm [23], $P2_1/m$ [21], C2/m [21], Pbnm [18], and $R\bar{3}c$ [22], have been proposed in the literature. Barring references [20] and [25], who have reported the existence of the γ-phase mainly on the basis of Raman scattering studies, the other workers could not capture the existence of the γ-phase in their structural studies.

In pure $BiFeO_3$, it is very difficult to identify unambiguously the crystal structure of the so-called β and γ phases at high temperatures, because $BiFeO_3$ is unstable at the reported α to β and β to γ transition temperatures and gradually decomposes into $Bi_2Fe_4O_9$ and $Bi_{25}FeO_{39}$/ $Bi_{25}FeO_{40}$ [20, 26, 27] followed by the onset of melting slightly above 1200K [20, 27]. The partial decomposition and off-stoichiometry of $BiFeO_3$ due to volatility of $Bi^{+3}$ at high temperatures close to the melting point seems to be largely responsible for the existing controversy about existence of the intermediate β-phase and the stability of the cubic γ-phase.

In order to capture unambiguously the sequence of structural phase transitions of $BiFeO_3$, we have recently undertaken a systematic study of structural phase transitions in several solid solutions of $BiFeO_3$ with $Pb(Fe_{1/2}Nb_{1/2})O_3$, $BaTiO_3$ [28] and $PbTiO_3$ [29], where the transition temperatures have been brought down well below the decomposition temperature of the perovskite phase in pure $BiFeO_3$. In this letter, we report results of high temperature x-ray diffraction (XRD) studies on $0.8BiFeO_3$-$0.2 Pb(Fe_{1/2}Nb_{1/2})O_3$(BF-0.2PFN). We have found the first unambiguous structural evidence for a paraelectric/paraelastic cubic phase (space group: $Pm\bar{3}m$) to which the ferroelectric/ferroelastic rhombohedral (space group: R3c) phase transforms without any decomposition problem. Contrary to the claim of previous workers [18, 20-23], we do not find any evidence for the so-called 'β' phase in stoichiometric monophasic perovskite composition of BF-0.2PFN. The R3c to $Pm\bar{3}m$ transition is shown to be of first order as revealed by the coexistence of two phases over a wide temperature range (ΔT~100K) and a discontinuous change in the unit cell volume.

The synthesis of phase pure samples of $BiFeO_3$ and its solid solutions in bulk continues to be a challenging task. One of the commonly reported impurity phases is $Bi_2Fe_4O_9$ [14]. Similarly the synthesis of phase pure $Pb(Fe_{1/2}Nb_{1/2})O_3$ is also very difficult due to easily formed pyrochlore phases such as $Pb_2Nb_2O_7$ and $Pb_2Nb_4O_{13}$ [30]. We have reduced the impurity phase content in our BF-0.2PFN samples to a nearly negligible level by optimizing the synthesis temperature and controlling the loss of $Bi_2O_3$ during sintering due to its significantly high vapour pressure at the sintering temperature. Analytic reagent (AR) grade chemicals, $Fe_2O_3$, $Nb_2O_5$, $PbCO_3$, and $Bi_2O_3$, with minimum assay of 99% or more, were used to synthesize BF-0.2PFN. Ball milled mixture of the ingredients was calcined in open atmosphere at 1133K. The calcined powder was pelletized using a steel die of 12 mm diameter in a uniaxial hydraulic press at an optimized load of 65 kN. A 2% polyvinyl alcohol (PVA) solution in water was used as a binder. The green pellets were kept at 773K for 10 h to burn off the binder material and then sintered at 1173K for 1 h in a sealed alumina crucible with calcined powder of the same composition used as a spacer powder to prevent the loss of $Bi_2O_3$ during sintering. There was no weight loss after sintering confirming the retention of the nominal composition even after sintering.

High temperature x-ray diffraction (XRD) measurements were carried out using an 18 kW rotating anode (Cu) based Rigaku powder diffractometer operating in the Bragg–Brentano geometry and fitted with a graphite monochromator in the diffracted beam. The data were collected in the 2θ range 20°–120° at a step of 0.02°. The XRD data were analyzed by Rietveld refinement technique using FULLPROF package. In the refinements, pseudo-Voigt function and a sixth order polynomial were used to define the profile shape and background, respectively. We have used the microstructural strain parameter and anisotropic thermal parameters similar to that in ref. [31] for pure $BiFeO_3$. The positional coordinates of atoms in the asymmetric unit of R3c space group of $BiFeO_3$ are (0,0,1/4+s), (0,0,t) and (1/6-2e-2d,1/3-4d,1/12) for $Bi^{3+}/Pb^{2+}$, $Fe^{3+}/Nb^{5+}$ and $O^{2-}$, respectively, using hexagonal unit cell [32], where the 's' and 't' parameters describe the polar cationic displacements, 'd' represents the octahedral distortion, and 'e' the angle of antiphase rotation $\Phi= \tan^{-1}4\sqrt{3}e$. Fig.1 shows the results of full pattern Rietveld refinement of BF-0.2PFN at room temperature for the R3c space group. The excellent fit between the observed and calculated profiles and the absence of any spurious peak due to impurities (see the inset), confirms that the structure of BF-0.2PFN is identical to that of pure $BiFeO_3$ at room temperature. The refined parameters are shown in Table 1. The values of s, t, d and e parameters of BF-0.2PFN are found to be nearly comparable to the values reported on $BiFeO_3$ using powder neutron diffraction data [31] with a slightly reduced tilt angle of $\Phi=11.7(4)°$ ( as against $12.8°$ for $BiFeO_3$).

Figure 2(a, b) shows the evolution of the 111 and 220 pseudocubic peaks (with respect to the elementary perovskite cell) from room temperature to 1023K. Both the peaks are doublets for the R3c space group. The splitting of the 111 and 220 peaks persists up to 973K and disappears in the profiles recorded at 998 and 1023K (see Fig. 2(c) and Fig. 2(d) for zoomed patterns at these temperatures). The antiphase rotation of the adjacent oxygen octahedra in the R3c space group leads to appearance of superlattice peaks, like 311 (w.r.t. doubled pseudocubic cell) near $2\theta=37.5°$ in Fig 2(a) and 2(c). This peak is discernible up to about 973K, as can be seen from Fig 2(c) and disappears at T≥998K. All these observations suggest that the α-phase of BF-0.2PFN in the R3c space group transforms to the cubic phase above 973K. However, the 220 pseudocubic profile shows a triplet structure in the temperature range 873 to 973K. The peak positions of the two outer peaks in Fig 2(b) match with the R3c peak positions while the inner peak position matches with the singlet peak position at T≥998K revealing the coexistence of the rhombohedral (R3c) and cubic (Pm$\bar{3}$m) phases over the temperature range 873K≤T<998K. This coexistence over ~100K temperature range reveals first order nature of the R3c to Pm$\bar{3}$m phase transition. We do not find evidence for any intermediate 'β' phase reported for BiFeO$_3$ in the literature [18, 20-23]. It is worth mentioning that Palai et al reported an orthorhombic 'β' phase on the basis of observation of triplet of peaks observed by them, which we believe could be due to the coexistence of R3c and Pm$\bar{3}$m phases, wrongly interpreted by these workers as an orthorhombic phase.

We now proceed to provide quantative confirmation of the foregoing observations by Rietveld analysis of the full powder diffraction profile in the phase coexistence region considering all the proposed structural models i.e., I4/mcm, Pbnm, C2/m, P2$_1$/m and R3c+Pm$\bar{3}$m space groups. Fig. 3 shows the observed, calculated and difference profiles of pseudocubic 111, 200, 210 and 220 reflections, obtained after full pattern refinement. Among the five possible structural models, the model considering coexistence of R3c and Pm$\bar{3}$m phases has the lowest number of refinable parameters and gives the best fit with lowest χ2 value, as can be seen from Fig. 3(e). The I4/mcm [23], Pbnm [18], C2/m [21], P2$_1$/m [21] space groups proposed in the literature may be rejected outrightly as they do not account for the observed peaks (see e.g. 111, 210 and 220 profiles in Fig. 3(a) to (d)). Since the phase coexistence region is quite wide (ΔT~100K), there exists the possibility of the ferroelectric/ferroelastic R3c phase having transformed into the R$\bar{3}$c phase within the two phase region, which also belongs to the a$^-$ a$^-$ a$^-$ tilt system but is paraelectric (because of the presence of the centre of symmetry). The results of full profile refinement for a possible R$\bar{3}$c phase coexisting with the Pm$\bar{3}$m phase are compared with R3c+Pm$\bar{3}$m model in Fig. 3(f) and (e). We used Hamilton criterion to test the statistical significance of the difference in $\chi^2$ for the R$\bar{3}$c+Pm$\bar{3}$m and R3c+Pm$\bar{3}$m structural models. We find that this criterion clearly favours the R3c+Pm$\bar{3}$m model. Rietveld refinement of x-ray diffraction pattern in the phase coexistence region at 873K, 898K, 923K, 948K and 973K gives the percentage of cubic phase as~ 7%, 9%, 10%, 40% and 75% before complete transformation to the cubic phase in the Pm$\bar{3}$m space group around 998K. Our results not only provide the first unambiguous evidence for the paraelectric/paraelastic cubic phase of $BiFeO_3$, but also reveal that the ferroelectric and antiferrodistortive (tilt) transitions occur at the same temperature leading to the non-centrosymmetric space group R3c with $a^- a^- a^-$ tilt system. This also implies that the soft R (q=1/2 1/2 1/2) and $\Gamma$ (q=0 0 0) point phonons of the cubic phase condense together in the R3c phase. Antiferrodistortive (tilt) transitions preceding ferroelectric instability are reported to frustrate the development of ferroelectric order in materials like $SrTiO_3$ [33] leading to the stabilization of the paraelectric phase down to 3K. BF-0.2PFN seems, however, to be unique in that the transition is driven simultaneously by two primary order parameters corresponding to $\Gamma 4^-$ and $R4^+$ irreps of the cubic space group.

Figure 4 depicts the variation of the hexagonal unit cell parameters and the elementary perovskite cell volume with temperature. It is interesting to note that the unit cell volume shows a small anomaly at $T_N$ (see the inset), similar to that in $BiFeO_3$ [21], $(Bi_{0.90} Ba_{0.10})(Fe_{0.90} Ti_{0.10})O_3$, and $(Bi_{0.73} Pb_{0.27})(Fe_{0.73} Ti_{0.27})O_3$ [15, 34] due to magnetoelastic coupling. More interestingly, there is a significant discontinuous change in the unit cell volume of the rhombohedral and cubic phases in the transition range confirming the first order character of the phase transition.

To summarize, we have presented first unambiguous evidence for the paraelectric phase of 0.8BF-0.2PFN without any decomposition and melting problem faced for pure $BiFeO_3$. The structure of the paraelectric/paraelastic phase is found to be cubic, and not tetragonal [23], orthorhombic [18, 20] or monoclinic [21] reported in the literature. The cubic to rhombohedral phase transition is of first order as confirmed by phase coexistence and discontinuous change of unit cell volume. We do not find any evidence of the so-called intermediate 'β' phase [20] of $BiFeO_3$. Our results suggest that both the R point and Γ point phonons freeze simultaneously to give the non-centrosymmetric space group R3c with antiphase rotated oxygen octahedra in the $a^- a^- a^-$ tilt system.

One of the authors (J. P. Patel) acknowledges support from UGC, India.

**Figure Captions**

**Figure1.** Observed (dots), calculated (continuous line) and difference (bottom curve) plots for BF-0.2PFN obtained after Rietveld refinement.

**Figure2.** Evolution of the x-ray diffraction profiles with temperature: (a) the 311 superlattice and 111 pseudocubic reflections, (b) the 220 pseudocubic reflection, (c) 311 superlattice reflection (zoomed pattern). (d) and (e) show the pseudocubic reflection 111 at 998 and 1023K. Asterisk in (a) indicates the 311 superlattice peak and the arrows in (b) indicate the cubic peak positions coexisting with the rhombohedral peaks.

**Figure3.** Observed (dots), calculated (line), and difference (bottom line) profiles for selected (111, 200, 210 and 220) pseudocubic reflections obtained after full pattern refinements using different space groups and structural models proposed in the literature and the present work. Vertical tick marks are the peaks position.

**Figure4.** Evolution of the R3c and the $Pm\bar{3}m$ cell parameters as a function of temperature. Inset shows small change in unit cell volume at the magnetic phase transition temperature.

**Table Caption**

**Table**1. Results of the Rietveld refinement of BF-0.2PFN at 300K.

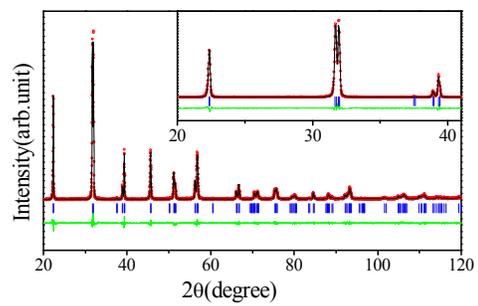

**Fig.1**

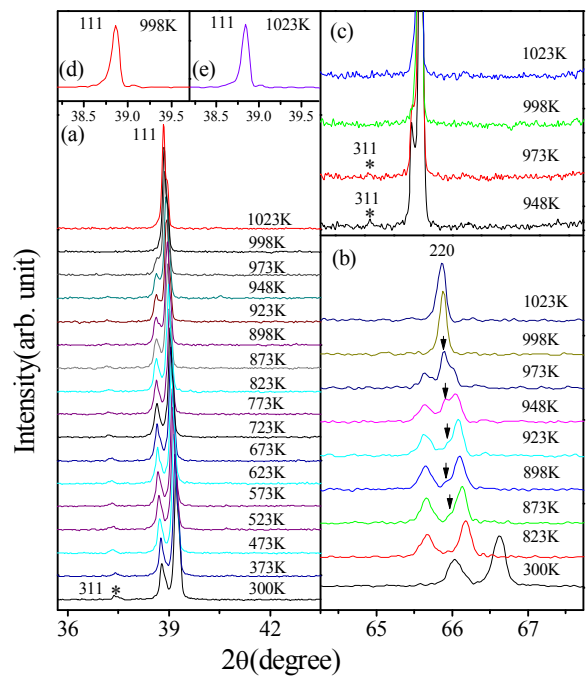

**Fig.2**

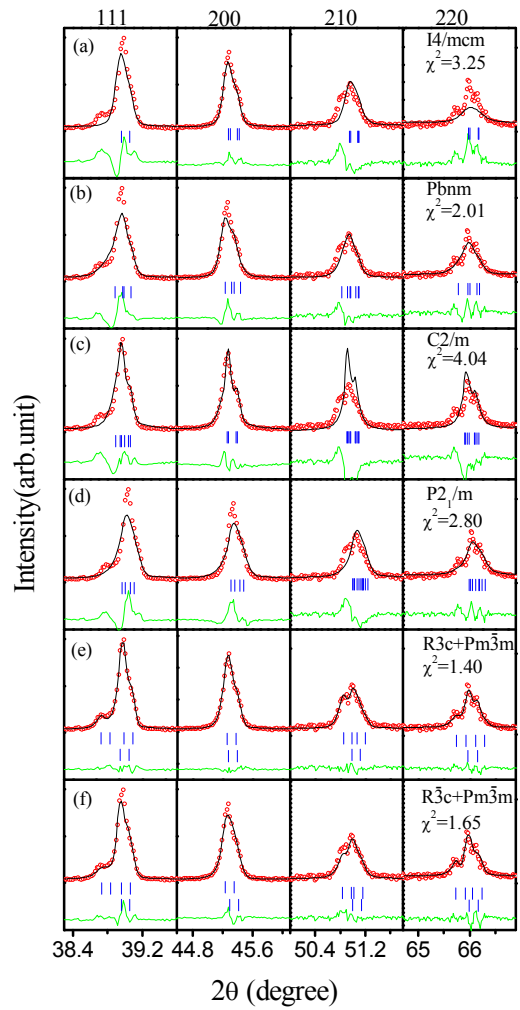

**Fig.3**

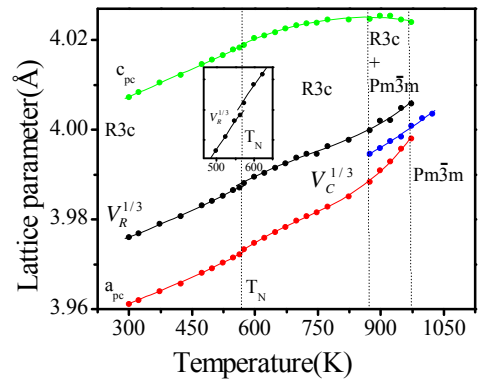

**Fig.4**

**Table 1**

| Ions | x | y | z | B (Å$^2$) |
|---|---|---|---|---|
| Bi$^{3+}$/Pb$^{2+}$ | 0.0 | 0.0 | 0.2943(7) | 0.98(2) |
| Fe$^{3+}$/Nb$^{5+}$ | 0.0 | 0.0 | 0.0193(9) | 0.012(8) |
| O$^{2-}$ | 0.221(2) | 0.345(2) | 0.08300 | 1.0(3) |
| a=b=5.6014(4) Å, c=13.8817(3) Å, α=β=90°, γ=120° | | | | |
| R$_p$=9.63, R$_{wp}$=14.3, R$_{exp}$=11.81, and χ$^2$=1.47 | | | | |